\documentclass[12pt]{revtex4} [showpacs,preprint,amsmath,amssymb]
\usepackage{graphicx}
\begin{document}
\title{Effect of triplet correlation on the equation of pair correlation function in a weakly coupled inhomogeneous plasma system}
\author{Anirban Bose}
\affiliation { Serampore College, Serampore, Hooghly, India.\\
}

\begin{abstract} It is observed that retaining the triplet correlation to derive the equation of pair correlation function from the first two members of BBGKY hierarchy, modifies the structure of the equation and the pair correlation function significantly. This equation may be used to explore the thermodynamic properties of the weakly coupled inhomogeneous plasma systems. This study may also be relevant for homogeneous plasmas.
\end{abstract}
\maketitle

\newpage
\section{Introduction}
Pair correlation function plays an important role to study the thermodynamic properties of plasma systems. Consequently, the determination and study of the pair correlation function remains a topic of interest even in present days \cite{kn:ve, kn:nr, kn:nd, kn:sd}. The Liouville equation provides an exact description of a plasma \cite{kn:ni, kn:ac}. However, since it is not practically possible to solve the equation exactly, we approximately solve it under certain physically acceptable conditions. In order to do that, we obtain an infinite coupled chain of integro-differential equations. These equations are called BBGKY hierarchy. In a previous article,
we have obtained an equation of pair correlation function from the first two members of BBGKY hierarchy \cite{kn:ab}. In another article, we have solved the equations  and obtained the pair correlation function of weakly coupled, weakly inhomogeneous plasma systems \cite{kn:ab1}. In the entire formalism we have neglected the triplet correlation function which is justified \cite{kn:on} for the plasma systems where the correlation is weak but not negligible. In this article, we have kept the triplet correlation function and solve the first two members of BBGKY hierarchy to obtain an equation of pair correlation function. This equation is capable of providing the thermodynamic properties of the plasma systems which we could not handle with our previous equation of pair correlation function. In fact, we have extended our formalism to be applicable to the plasma systems with higher range of correlation.

\section{Derivation of the equation of pair correlation in thermal equilibrium}

We consider an inhomogeneous dusty plasma system. The dusts are weakly correlated with the ions and electrons are uncorrelated. We assume, for simplicity, that all the  dust particles have same mass m and same charge q. Under these assumptions, the first two members of BBGKY hierarchy for the dust particles, are written as 

\begin{equation} \frac{\partial f_1}{\partial t} + {\bf v_1}\cdot\frac{\partial f_1}{\partial {\bf x_1}}
+ n_0\int d{\bf X_2} {\bf a_{12}}\cdot\frac{\partial}{\partial{\bf
v_1}}\left [ f_1({\bf X_1})f_1({\bf X_2})+g_{12}\right ] = 0,
\label{v1}\end{equation}
 \begin{eqnarray}
&&\frac{\partial g_{12}}{\partial t } + {\bf v_{1}}\cdot
\frac{\partial g_{12}}{\partial {\bf x_{1}}}+{\bf v_{2}}\cdot
\frac{\partial g_{12}}{\partial {\bf x_{2}}}+n_{0}\int d{\bf
X_{3}}f_{1}({\bf X_3})\left( {\bf a_{13}}\cdot \frac{\partial
g_{12}}{\partial {\bf v_1}}+ {\bf a_{23}}\cdot\frac{\partial
g_{12}}{\partial {\bf v_2}}\right ) \nonumber\\
 & &=-\left ({\bf a_{12}}\cdot\frac{\partial}{\partial
{\bf v_{1}}}+{\bf a_{21}}\cdot\frac{\partial}{\partial
{\bf v_{2}}}\right )\left[f_{1}({\bf X_1})f_{1}({\bf X_2})+g_{12}\right]\nonumber\\
& &-n_{0}\int d{\bf X_{3}}\left [{\bf a_{13}}\cdot\frac{\partial
f_{1}({\bf X_1})}{\partial {\bf v_{1}}}g_{23}+{\bf
a_{23}}\cdot\frac{\partial f_{1}({\bf X_2})}{\partial {\bf
v_{2}}}g_{13}\right ]\nonumber\\
& &-n_{0}\int d{\bf X_{3}}\left [{\bf a_{13}}\cdot\frac{\partial
h_{123}}{\partial {\bf v_{1}}}+{\bf
a_{23}}\cdot\frac{\partial h_{123}}{\partial {\bf
v_{2}}}\right ]\label{v2}\end{eqnarray} where ${\bf X}={\bf (x,v)}$, $f_{1}$ is the single particle distribution, $g_{12}$ is the pair correlation function
and
$${\bf a_{ij}}=-\frac{1}{m}\frac{\partial}{\partial {\bf x_i}}{\phi_{ij}},$$
where $${\phi_{ij}}=-\frac{q^2}{|\bf{x_{i}}- x_{j}|}.$$

$a_{ij}$ denotes acceleration of the i th particle due
to force exerted by the j th particle and $\phi_{ij}$ is the
energy of interaction between them.

The system is considered to be in thermal equilibrium. Hence, the first terms of eqs.({\ref{v1}}) and ({\ref{v2}}) are ignored. The pair correlation function ($g_{12}$) is written as
\begin{eqnarray}
 g_{12}({\bf {X_1,X_2}})&=&f_{1}({\bf X_1})f_{1}({\bf X_2})\chi_{12}({\bf x_1,x_2}).
\label{v3}\end{eqnarray}
The triplet correlation function ($h_{123}$) is written as \cite{kn:on}
\begin{eqnarray}
 h_{123}({\bf {X_1,X_2,X_3}})&=&f_{1}({\bf X_1})f_{1}({\bf X_2})f_{1}({\bf X_3})[\chi_{12}({\bf x_1,x_2})\chi_{13}({\bf x_1,x_3})\nonumber\\
 & &+\chi_{12}({\bf x_1,x_2})\chi_{23}({\bf x_2,x_3})\nonumber\\
 & &+\chi_{13}({\bf x_1,x_3})\chi_{23}({\bf x_2,x_3})\nonumber\\
 & & +\int d{\bf X_{4}}f_{1}({\bf X_4})\chi_{14}({\bf x_1,x_4})\chi_{24}({\bf x_2,x_4})\chi_{34}({\bf x_3,x_4})].
\label{v31}\end{eqnarray}
Single particle distributions ($f_{1}({\bf X_1})$, $ f_{1}({\bf X_2})$ and $ f_{1}({\bf X_3})$) are functions of position and
velocity. $\chi_{ij}$ is a symmetric function of $\bf x_i$
and $\bf x_j$. Using eqs.({\ref{v1}}) and ({\ref{v3}}), we obtain
\begin{eqnarray}
 {\bf v_{2}}\cdot\frac{\partial g_{12}}{\partial
{\bf x_{2}}}&=&f_{1}({\bf X_1})\chi_{12} {\bf
v_{2}}\cdot\frac{\partial
f_{1}({\bf X_2})}{\partial {\bf x_{2}}}+f_{1}({\bf X_1})f_{1}({\bf X_2}){\bf v_{2}}\cdot\frac{\partial\chi_{12}}{\partial{\bf x_{2}}},\label{v41}\\
{\bf v_{2}}\cdot\frac{\partial f_{1}({\bf
X_2})}{\partial\bf{x_{2}}} &=& - n_{0}\int d{\bf X_{3}} {\bf
a_{23}}\cdot \frac{\partial}{\partial {\bf v_{2}}} (f_{1}({\bf
X_2})f_{1}({\bf X_3})+g_{23}),
\label{v4}\end{eqnarray}

and similar expressions for ${\bf v_{1}}\cdot{\partial
g_{12}}/{\partial {\bf x_{1}}}$. Using Eqs.({\ref{v41}}) and ({\ref{v4}}) in eq.({\ref{v2}}), we obtain

\begin{eqnarray}
&&f_{1}({\bf X_1})f_{1}({\bf X_2})\left [{\bf
v_{1}}\cdot\frac{\partial \chi_{12}}{\partial{\bf x_{1}}}+{\bf
v_{2}}\cdot\frac{\partial \chi_{12}}{\partial{\bf x_{2}}} \right ]
\nonumber\\
& &-n_{0}\int d{\bf X_{3}}\left[f_{1}({\bf X_2})\chi_{12}{\bf a_{13}}\cdot \frac{\partial
g_{13}}{\partial {\bf v_{1}}}+f_{1}({\bf X_1})\chi_{21}{\bf a_{23}}\cdot
\frac{\partial g_{23}}{\partial {\bf v_{2}}}\right
]\nonumber\\
& &=-\left ({\bf a_{12}}\cdot\frac{\partial}{\partial {\bf
v_{1}}}+{\bf a_{21}}\cdot\frac{\partial}{\partial
{\bf v_{2}}}\right )\left (f_{1}( {\bf X_1})f_{1}({\bf X_2})+g_{12} \right )\nonumber \\
& &-n_{0}\int d{\bf X_{3}}\left [{\bf a_{13}}\cdot \frac{\partial
f_{1}({\bf X_1})}{\partial {\bf v_{1}}}g_{23}+{\bf a_{23}}\cdot
\frac{\partial f_{1}({\bf X_2})}{\partial {\bf v_{2}}}g_{13}\right
]\nonumber\\
& &-n_{0}\int d{\bf X_{3}}\left [{\bf a_{13}}\cdot\frac{\partial
h_{123}}{\partial {\bf v_{1}}}+{\bf
a_{23}}\cdot\frac{\partial h_{123}}{\partial {\bf
v_{2}}}\right ].
\label{v5}\end{eqnarray}

The single
particle distribution functions are written in the following form
$$f_1 ({\bf X_1}) = f_M({\bf v_1})F_1({\bf x_1}),$$
where $f_M$ is a Maxwellian distribution and $F_1$ is the space part.

Using eqs.({\ref{v3}}) and ({\ref{v31}}) in eq.({\ref{v5}}) and following the method as described in \cite{kn:ab},  we obtain

\begin{eqnarray}
&&\frac{\partial \chi_{12}}{\partial{\bf x_{1}}}
+\frac{1}{k_{B}T}\frac{\partial \phi_{12}•}{\partial{\bf x_1}•}+\frac{1}{k_{B}T}\frac{\partial \phi_{12}•}{\partial{\bf x_1}•}\chi_{12}
+ \frac{n_{0}}{k_{B}T}\int d{\bf {X_{3}}}\frac{\partial{\phi_{13}}}{{\partial {\bf x_1}}} f_{1}({\bf X_3}) \chi_{23}\nonumber\\
 & & +\frac{n_{0}\chi_{12}}{k_{B}T}\int d{\bf {X_{3}}}\frac{\partial{\phi_{13}}}{{\partial {\bf x_1}}} f_{1}({\bf X_3}) \chi_{23}+\frac{n_{0}}{k_{B}T}\int d{\bf {X_{3}}}\frac{\partial{\phi_{13}}}{{\partial {\bf x_1}}} f_{1}({\bf X_3})\chi_{13}\chi_{23}\nonumber\\
 & &+\frac{n_{0}}{k_{B}T}\int d{\bf {X_{3}}}d{\bf {X_{4}}}\frac{\partial{\phi_{13}}}{{\partial {\bf x_1}}} f_{1}({\bf X_3})f_{1}({\bf X_4})\chi_{14}\chi_{24}\chi_{34} =0.\label{v7111}\end{eqnarray}

This is the equation of the pair correlation function, which is derived from the BBGKY hierarchy. Previously, we neglected the triplet correlation function to obtain the equation of pair correlation function \cite{kn:ab}. Therefore, we observe some additional terms in eq.({\ref{v7111}}), which are not present in the previous equation (eq.(11) of \cite{kn:ab}). We may apply this equation to both homogeneous and inhomogeneous systems and it is capable of describing both short and long range behaviours. 

In order to derive this complex equation, we have chosen a specific form of the triplet correlation function. This form is previously obtained by O'neil and Rostoker \cite{kn:on} for the homogeneous plasma systems.

It is not obvious that the form of the triplet correlation function obtained by O'neil and Rostoker \cite{kn:on} for the homogeneous system may be applied to the inhomogeneous systems. 

However, we may guess the triplet function of the inhomogeneous system, which must obey the following properties:

(1) This function, for both homogeneous and inhomogeneous cases, must be invariant with respect to interchange of particle indices. 

(2) It must reduce to the form derived by O'neil and Rostoker \cite{kn:on} in the homogeneous limit.

(3) This function should not be higher order than $g^2$.

We may construct a particular form of the triplet correlation function, consistent with the above conditions. 

\begin{eqnarray}
 h_{123}({\bf {X_1,X_2,X_3}})&=&f_{1}({\bf X_1})f_{1}({\bf X_2})f_{1}({\bf X_3})[\chi_{12}({\bf x_1,x_2})\chi_{13}({\bf x_1,x_3})\chi_{23}({\bf x_2,x_3})\nonumber\\
 & &+\chi_{12}({\bf x_1,x_2})\chi_{13}({\bf x_1,x_3})\nonumber\\
 & &+\chi_{12}({\bf x_1,x_2})\chi_{23}({\bf x_2,x_3})\nonumber\\
 & &+\chi_{13}({\bf x_1,x_3})\chi_{23}({\bf x_2,x_3})\nonumber\\
 & & +\int d{\bf X_{4}}f_{1}({\bf X_4})\chi_{14}({\bf x_1,x_4})\chi_{24}({\bf x_2,x_4})\chi_{34}({\bf x_3,x_4})].
\label{v931}\end{eqnarray}

This function may be assumed to be the triplet correlation function. Certainly, the expressions of the pair correlation functions ($\chi_{ij}$), with which the triplet function is constructed, are modified in the presence of inhomogeneity, but the symmetric structure of the triplet correlation function should be maintained. It is necessary to discuss the properties of the function in different limits, and check if it is physically consistent.

If the particles in the triplet are close to each other (separations are of the order of Landau length), the first four terms in eq.(\ref{v931}) are important and the last term may be dropped. This is true because, due to close proximities, the terms coming from the discrete effects are stronger than the terms from the collective effects. Similarly, if the particles in the triplet are separated from each other by distances of the order of the Debye length, the last four terms are more important and the first term may be dropped. These results, in the homogeneous limit, are consistent with the results obtained by O'neil and Rostoker \cite{kn:on}.

Finally, if we move away from both the above limits to reach the mid range, and match the triplet correlation, we may find the three terms in the middle are more important in comparison to the first and last terms. 

If we had replaced eq.(\ref{v31}) with the triplet correlation function given by eq.(\ref{v931}), it  would have produced an additional term to eq.({\ref{v7111}}). The new equation, whose fifth term is the additional term, is given below. 

\begin{eqnarray}
&&\frac{\partial \chi_{12}}{\partial{\bf x_{1}}}
+\frac{1}{k_{B}T}\frac{\partial \phi_{12}•}{\partial{\bf x_1}•}+\frac{1}{k_{B}T}\frac{\partial \phi_{12}•}{\partial{\bf x_1}•}\chi_{12}
+ \frac{n_{0}}{k_{B}T}\int d{\bf {X_{3}}}\frac{\partial{\phi_{13}}}{{\partial {\bf x_1}}} f_{1}({\bf X_3}) \chi_{23}\nonumber\\
& & +\frac{n_{0}\chi_{12}}{k_{B}T}\int d{\bf {X_{3}}}\frac{\partial{\phi_{13}}}{{\partial {\bf x_1}}} f_{1}({\bf X_3})\chi_{23}\chi_{13}\nonumber\\ 
& & +\frac{n_{0}\chi_{12}}{k_{B}T}\int d{\bf {X_{3}}}\frac{\partial{\phi_{13}}}{{\partial {\bf x_1}}} f_{1}({\bf X_3}) \chi_{23}+\frac{n_{0}}{k_{B}T}\int d{\bf {X_{3}}}\frac{\partial{\phi_{13}}}{{\partial {\bf x_1}}} f_{1}({\bf X_3})\chi_{13}\chi_{23}\nonumber\\
 & &+\frac{n_{0}}{k_{B}T}\int d{\bf {X_{3}}}d{\bf {X_{4}}}\frac{\partial{\phi_{13}}}{{\partial {\bf x_1}}} f_{1}({\bf X_3})f_{1}({\bf X_4})\chi_{14}\chi_{24}\chi_{34} =0.\label{v71111}\end{eqnarray}

\section{Results and Discussions}
In previous section, we have derived an equation of pair correlation function  of a weakly coupled inhomogeneous plasma system from BBGKY hierarchy. 

In order to obtain the equation, we have assumed that pair and triplet correlation functions are the product of the single particle distribution functions and some  other functions which are symmetric with respect to the positions of the concerned particles.  

As discussed by O'neil and Rostoler \cite{kn:on}, we may identify the long and short range regions of eq.({\ref{v7111}}). First of all, the relative importance of the different terms of eq.({\ref{v7111}}) may be investigated.
In order to do that, we introduce a small parameter $g=1/n\lambda_{D}^3$, where n is the average density and $\lambda_{D}$ is the Debye length.
The different terms of eq.({\ref{v7111}}) have the following order of magnitude.

$$1:\frac{\phi_{12}}{k_{B}T}\frac{1}{\chi_{12}}:\frac{\phi_{12}}{k_{B}T}:\frac{<\phi >}{k_{B}T}\frac{\chi}{\chi_{12}}(n\lambda_{D}^3):\frac{<\phi>}{k_{B}T}\chi \frac{\chi_{12}}{\chi_{12}}(n\lambda_{D}^3):\frac{<\phi>}{k_{B}T}\frac{\chi^{2}}{\chi_{12}}(n\lambda_{D}^3)$$
\begin{equation}:\frac{<\phi>}{k_{B}T}\frac{\chi^{3}}{\chi_{12}}(n\lambda_{D}^3)^{2}\label{v991}\end{equation}
In this connection, we have deliberately kept the indices of particle number one and two in the pair correlation term ($\chi_{12}$). This enables us to control the distance between particle one and two, and investigate the properties of pair correlation function for different ranges of separation.

$\chi$ and $<\phi>/k_{B}T$ are of the order of $g$.
To explore the asymptotic limit, we set the distance between particle one and particle two to be of the order of Debye length. In that limit, $\chi_{12}$ is also of the order of $g$.
Therefore, neglecting terms of higher order than $g$, the equation of pair correlation in the asymptotic limit is given by:

\begin{eqnarray}
\frac{\partial \chi_{12}}{\partial{\bf x_{1}}}
+ \frac{n_{0}}{k_{B}T}\int d{\bf {X_{3}}}\frac{\partial{\phi_{13}}}{{\partial {\bf x_1}}} f_{1}({\bf X_3}) \chi_{23}
=-\frac{1}{k_{B}T}\frac{\partial \phi_{12}}{\partial{\bf x_1}}.\label{v71112}\end{eqnarray}

To look into the short range region, we set the distance between the pairing particles (particle one and two) to be of the order of Landau length, and the equation of pair correlation function is given by: 
\begin{eqnarray}
\frac{\partial \chi_{12}}{\partial{\bf x_{1}}}
+ \frac{n_{0}}{k_{B}T}(1+\chi_{12})\int d{\bf {X_{3}}}\frac{\partial{\phi_{13}}}{{\partial {\bf x_1}}} f_{1}({\bf X_3}) \chi_{23}
=-\frac{1}{k_{B}T}\frac{\partial \phi_{12}}{\partial{\bf x_1}}(1+\chi_{12}).\label{v71110}\end{eqnarray}

The equation of pair correlation is a complex equation. Therefore, we need to simplify eq.({\ref{v7111}}) under proper physical conditions. For example, we may gradually decrease the distance between particle one and two so that the relative importance of the first term of the last three terms of eq.({\ref{v991}}) is increased over the other two, and when the distance is of the order of Landau length, the relative strength of the terms of equation of pair correlation function is given by  
$$1:1:1:g:g:g^{2}:g^{2}$$.
Hence, we may drop the last two terms of eq.({\ref{v991}}).
If, instead of eq.(\ref{v31}), we had started with the other expression of triplet correlation function (eq.(\ref{v931})), we would have obtained the same result, since the contribution of the first term of the triplet correlation of eq.(\ref{v931}) would be of the order of $g^2$.

\begin{eqnarray}
&&\frac{\partial \chi_{12}}{\partial{\bf x_{1}}}
+\frac{1}{k_{B}T}\frac{\partial \phi_{12}•}{\partial{\bf x_1}•}+\frac{1}{k_{B}T}\frac{\partial \phi_{12}•}{\partial{\bf x_1}•}\chi_{12}
+ \frac{n_{0}}{k_{B}T}\int d{\bf {X_{3}}}\frac{\partial{\phi_{13}}}{{\partial {\bf x_1}}} f_{1}({\bf X_3}) \chi_{23}\nonumber\\
 & & +\frac{n_{0}\chi_{12}}{k_{B}T}\int d{\bf {X_{3}}}\frac{\partial{\phi_{13}}}{{\partial {\bf x_1}}} f_{1}({\bf X_3}) \chi_{23}=0
.\label{v71}\end{eqnarray}
The solution of this equation is 
\begin{eqnarray}
\chi_{12}=e^{-\psi_{12}/k_{B}T}-1,\label{v711}\end{eqnarray}
where

\begin{eqnarray}
\psi_{12}
=\phi_{12}
+n_{0}\int d{\bf {X_{3}}}\phi_{13} f_{1}({\bf X_3}) \chi_{23}.
\label{v81}\end{eqnarray}

We may compare the above results with what we already have in literature. For example, in the short range behaviour, we have \cite{kn:akhi}
\begin{eqnarray}
\chi_{12}=e^{-\phi_{12}/k_{B}T}-1,\label{v712}\end{eqnarray}
which differs from the expression given by eq.({\ref{v711}}). If we bring the pairing particles closer, we may neglect the second term of eq.({\ref{v81}}) in comparison to the first term and the two expressions given by eq.({\ref{v711}}) and eq.({\ref{v712}}) become identical. On the other hand, if we increase the distance between the particles, the second term will gradually become comparable with the first term. In that situation, the pair correlation in eq.({\ref{v711}}) is more appropriate than what is given by eq.({\ref{v712}}). In \cite{kn:akhi}, the authors have obtained an interpolation formula by adding a factor of $e^{-r\lambda_{D}}$ under the exponential sign and claimed that this form is applicable for all distances between the pairing particles. In our calculation, without introducing any ad hoc term,  we have shown it rigorously that the potential under the exponential will be modified by the presence of collective effects (second term of eq.({\ref{v711}})). In order to do that, we need to keep the triplet correlation term in the second member of BBGKY hierarchy. In \cite{kn:fri}, the authors have obtained a different expression of pair correlation function, which is valid for all distances. However, the formalism does not not include the effect of triplet  correlation function and is restricted to homogeneous plasma systems. If we drop the third and last term of eq.({\ref{v71}}), we get the equation previously derived by Chavanis  from a  BBGKY-like hierarchy for inhomogeneous and homogeneous systems with long range interactions \cite{kn:ph5, kn:ph6}.

\section{Conclusion}
In this article we have explored how the presence of triplet correlation modifies the equation of pair correlation function derived from the first two members of BBGKY hierarchy. However, we have chosen a particular form of the triplet correlation function to obtain the final equation of pair correlation. In fact, in my previous article \cite{kn:ab}, we have obtained few terms which are of the same order of those terms which could have been originated from the triplet correlation function in the second member of BBGKY hierarchy. Hence, for the sake of completeness, we should have retained the triplet correlation function in the second member of BBGKY hierarchy. We have chosen a specific form of the triplet correlation function of the inhomogeneous plasma system. This particular form was observed earlier by O'neil and Rostoker \cite{kn:akhi} in the homogeneous system. Finally, we have been able to provide a formal solution of the pair correlation under the condition that the particles in the pair are not very far away from each other. This pair correlation exhibits both the short and long range effects. Inclusion of triplet correlation gives the opportunity to explore the plasma systems where the strength of correlation is higher than what we could have done without the triplet term.

At the end, apart from the inhomogeneous cases, this study may also be relevant for homogeneous plasmas. Usually, while calculating the pair correlation function, the triplet term is dropped in the equation of pair correlation function at the very beginning. In this article our aim is to keep the triplet term in the calculation and see how the pair correlation function gets modified in its presence.

This research was supported by the Department of Science $\&$ Techonology and Biotechnology, West Bengal.


\newpage

\begin{thebibliography}{99}
\bibitem{kn:ve} V. E. Fortov, A. V. Gavrikov, O. F. Petrov, I. A. Shakhova, and V. S. Vorobev, Phys. Plasmas {\bf{14}}, 040705 (2007).
\bibitem{kn:nr} N. R. Shaffer, S. K. Tiwari, and S. D. Baalrud, Phys. Plasmas {\bf{24}}, 092703 (2017).
\bibitem{kn:nd} N. Desbiens, P. Arnault, and J. Clerouin, Phys. Plasmas {\bf{23}}, 092120 (2016).
\bibitem{kn:sd} S. D. Baalrud and J. Daligault, Phys. Plasmas {\bf{26}}, 082106 (2019).
\bibitem{kn:ni} D. R. Nicholson, Introduction to Plasma Theory (New York, Wiely, 1983).
\bibitem{kn:ac} A. Campa, T. Dauxois, D. Fanelli and S. Ruffo, Physics of long-range interacting systems (Oxford, Oxford University Press 2014).
\bibitem{kn:fb3} F. Bouchet, S. Gupta, and D. Mukamel, Physica A: Statistical Mechanics and its Applications {\bf{389}}, 4389 (2010).
\bibitem{kn:ab} A. Bose, Phys. Plasmas {\bf{23}}, 104505 (2016).
\bibitem{kn:ab1} A. Bose, Phys. Plasmas {\bf{26}}, 064501 (2019).
\bibitem{kn:on} T. ONeil and N. Rostoker, Phys. Fluids {\bf{8}}, 1109 (1965).
\bibitem{kn:akhi} A. Akhiezer, I. Akhiezer, R. Polovin, A. Sitenko and K. Stepanov, Plasma Electrodynamics. Vol. I: Linear Theory (Oxford–New York, Pergamon Press, 1975).
\bibitem{kn:fri} E. A. Frieman and D. L. Book, Phys. Fluids {\bf{6}}, 1700 (1963).
\bibitem{kn:ph5} P. H. Chavanis, Physica A, {\bf{387}}, 5716 (2008).
\bibitem{kn:ph6} P. H. Chavanis, Physica A, {\bf{361}}, 55 (2006).
\end{thebibliography}
\end{document}